\documentclass[
aps,
nofootinbib,
prd,
superscriptaddress,
tightenlines,
notitlepage,
twocolumn,
showpacs,
floatfix
]{revtex4-1}
\usepackage{amsmath}
\usepackage{latexsym}
\usepackage{amsfonts}
\usepackage{graphicx}
\usepackage{mathrsfs}
\usepackage{epstopdf}
\usepackage{subfigure}
\usepackage[utf8]{inputenc}
\usepackage{CJK}
\usepackage{float}
\usepackage{xcolor}
\usepackage{color}
\usepackage{hyperref}
\usepackage{diagbox}
\usepackage{bm}
\usepackage{lipsum}
\usepackage[normalem]{ulem}
\hypersetup{
colorlinks=true,
linkcolor=red,
citecolor=blue,
}

\usepackage{enumerate}

\newcommand{\reply}[1]{#1}

\begin{document}

\title{\reply{The Impact of Spin in} Compact Binary Foreground Subtraction for Estimating the Residual Stochastic Gravitational-wave Background in Ground-based Detectors}

\author{Hanlin Song}
\affiliation{School of Physics, Peking University, Beijing 100871, China}

\author{Dicong Liang}
\affiliation{Kavli Institute for Astronomy and Astrophysics, Peking University,
Beijing 100871, China}
\affiliation{Department of Mathematics and Physics, School of Biomedical Engineering, Southern Medical University, Guangzhou 510515, China}

\author{Ziming Wang}
\affiliation{Kavli Institute for Astronomy and Astrophysics, Peking University,
Beijing 100871, China}

\author{Lijing Shao}\email[Corresponding author: ]{lshao@pku.edu.cn}
\affiliation{Kavli Institute for Astronomy and Astrophysics, Peking University,
Beijing 100871, China}
\affiliation{National Astronomical Observatories, Chinese Academy of Sciences,
Beijing 100012, China}

\begin{abstract}
Stochastic gravitational-wave (GW) background (SGWB) contains information about
the early Universe and astrophysical processes. The recent evidence of SGWB by
pulsar timing arrays in the nanohertz band is a breakthrough in the GW
astronomy.  For ground-based GW detectors,  while 
in data analysis,
the SGWB can be masked by loud GW events from compact binary coalescences
(CBCs).  Assuming a next-generation ground-based GW detector network, we
investigate the potential for detecting the astrophysical and cosmological SGWB
with non-CBC origins by subtracting recovered foreground signals of loud CBC
events.  The Fisher Information Matrix (FIM) method is adopted for quick calculation.  As an extension of the studies by Sachdev {\it et al.} (2020) and Zhou
{\it et al.} (2023), two more essential features are considered. 
 Firstly, we incorporate non-zero aligned or anti-aligned spin parameters in our waveform
model.  Because of the inclusion of spins, we obtain significantly more
pessimistic results than the previous work, where the residual energy density of
foreground is even larger than the original CBC foreground. 
 \reply{For the most extreme case, we observe that the subtraction results are approximately 10 times worse for binary black hole events and 20 times worse for binary neutron star events than the scenarios without accounting for spins.}
The degeneracy between
the spin parameters and \reply{the} symmetric mass ratio is strong in the parameter
estimation process, and it contributes most to the imperfect foreground
subtraction.  Secondly, in this work, extreme CBC events with condition numbers of FIMs $c_{\rm{\Gamma}}>10^{15}$ are preserved. 
 The impacts of these extreme events on foreground subtraction are discussed.  Our results have important implications for assessing the
detectability of SGWB from non-CBC origins for ground-based GW detectors.
\end{abstract}
\maketitle

\allowdisplaybreaks 

\section{Introduction}

Recently, intriguing \reply{evidence of the Hellings-Downs} correlation
\cite{Hellings:1983fr} for gravitational-wave (GW) signals in the nanohertz
band were revealed by several pulsar timing arrays (PTAs),  including the North
American Nanohertz Observatory for Gravitational waves (NANOGrav)
\cite{NANOGrav:2023hde, NANOGrav:2023gor}, the European PTA (EPTA), along with
the Indian PTA (InPTA) \cite{EPTA:2023sfo, EPTA:2023akd, EPTA:2023fyk}, the
Parkes PTA (PPTA) \cite{Zic:2023gta, Reardon:2023gzh}, and the Chinese PTA
(CPTA) \cite{Xu:2023wog}.  These signals might have originated from the
stochastic GW background (SGWB).  Broadly speaking, the SGWB may arise from
multiple origins, including cosmological and astrophysical phenomena
\cite{Christensen:2018iqi}.  Possible cosmological origins include inflation
\cite{Turner:1996ck}, cosmic strings \cite{Siemens:2006yp, Damour:2004kw},
first-order phase transitions \cite{ Caprini:2007xq, Kosowsky:1991ua,
Huber:2008hg, Li:2021qer, Chen:2023bms}, and so on.  Astrophysical origins
include asymmetry of supernovae \cite{Fryer:2011zz},  core collapse of
supernovae \cite{Crocker:2017agi}, cumulative effects of compact binary
coalescences (CBCs) \cite{Cutler:2005qq, Zhu:2012xw, Zhu:2011bd, Phinney:2001di,
Rosado:2011kv}, and so on.

At present, the ground-based GW detector network is searching for signals from both astrophysical and cosmological SGWB in the audio frequency band.  The cross-correlation method is adopted, which assumes the SGWB is correlated between detectors while instrument noise is not correlated \cite{LIGOScientific:2009qal, LIGOScientific:2016jlg}. 
After processing the data from the three observing runs (O1, O2, and O3) of LIGO/Virgo/KAGRA detectors, no evidence for SGWB was found \cite{KAGRA:2021kbb}.
To detect a persistent signal from the SGWB, one needs a
long-term observation to improve the sensitivity.  However, during such an
observation, a considerable number of CBC events form a loud foreground,
weakening the ability to detect the SGWB from other origins. 
 Currently, nearly
one hundred CBC events are published in the first three observing runs of the
LIGO/Virgo/KAGRA (LVK) collaboration \cite{LIGOScientific:2021djp,
LIGOScientific:2018mvr, LIGOScientific:2020ibl, LIGOScientific:2021usb,
Nitz:2021zwj, Olsen:2022pin}.  In the future, with \reply{the} deployment of the
next-generation (XG) ground-based detectors, including the Einstein Telescope
(ET) \cite{Punturo:2010zz, Sathyaprakash:2019yqt} and the Cosmic Explorer (CE)
\cite{Reitze:2019iox, Kalogera:2021bya}, thousands of CBC events will be
detected annually \cite{Borhanian:2022czq, Ronchini:2022gwk, Iacovelli:2022bbs}.
In this work, we are interested in exploring the prospect of observing the non-CBC origin SGWB with the XG ground-based GW detector network.
Therefore, one needs to carefully deal with this
foreground composed of CBC events \cite{Cutler:2005qq, Zhu:2012xw,
Regimbau:2016ike, Sachdev:2020bkk, Zhou:2022nmt, Zhong:2022ylh, Pan:2023naq,
Bellie:2023jlq}.

Previously, \citet{Sachdev:2020bkk} considered the detection of non-CBC origin SGWB in a
network of XG detectors.  They adopted the Fisher Information Matrix (FIM) method
to quickly estimate the residual background after subtracting the resolved CBC
events.  For both binary black hole (BBH) and binary neutron star (BNS) events,
\citet{Sachdev:2020bkk} used a post-Newtonian (PN) expansion waveform with only
3 free binary parameters, i.e.\ the coalescence time $t_c$, the coalescence
phace $\phi_c$, and the chirp mass $\mathcal{M}$.  They found that the residual
background from BNS events is too large, limiting the capability of observing
SGWB from non-CBC origins, while the signals from BBH events can be subtracted
sufficiently such that their effect is negligible.  However, a recent study by
\citet{Zhou:2022nmt} showed pessimistic results.  They adopted the same method
as \citet{Sachdev:2020bkk} but added another 6 free parameters that are normal
in real parameter estimation  (PE\footnote{Here in this work, we use PE to mean that, when the signal-to-noise ratio of an event is large enough, the FIM method is adopted to obtain a multivariate Gaussian distribution to mimic the posterior.  Afterwards we use this posterior to draw samples.  It is distinct from the normal PE studies in real GW data.}) of CBCs, including the symmetric mass ratio
$\eta$, the redshift $z$, the right ascension $\alpha$, the declination
$\delta$, the orbital inclination angle $\iota$, and the GW polarization angle
$\psi$.  For simplicity, \citet{Zhou:2022nmt} have set the spins to zero for all
CBC events and adopted the \texttt{IMRPhenomC}  and \texttt{IMRPhenomD} models
to generate waveforms.  They found that including more parameters leads to a
significantly larger residual background for both BBH and BNS events than what
was found by~\citet{Sachdev:2020bkk}.  This is mainly due to the degeneracy
between the luminosity distance $D_L$ and the orbital inclination angle $\iota$,
as well as the degeneracy between the coalescence phase $\phi_c$ and the
polarization angle $\psi$.  There are also other methods for further foreground subtraction, such as \citet{Cutler:2005qq} and \citet{Pan:2023naq}. For the method of measuring the SGWB without subtracting the foreground, readers are referred to  \citet{Biscoveanu:2020gds}. 

In this work, we consider two more essential features built upon the subtraction methods in \citet{Sachdev:2020bkk} and \citet{Zhou:2022nmt}.  Firstly, the non-zero spins aligned or anti-aligned with the orbital angular momentum in the waveform are considered.  Common envelope evolution theory predicts that BBH events have nearly aligned spins \cite{Kalogera:1999tq, Mandel:2009nx, Dominik:2013tma, Eldridge:2017cyw, Giacobbo:2017qhh, Olejak:2020oel, LIGOScientific:2020kqk}. 
Meanwhile, the majority of observed BBH events in GWTC-3 \reply{preferentially have} aligned spins, and evidence of events with anti-aligned spins is also observed \cite{KAGRA:2021duu}.
For BNS events, the spin axis will become aligned with orbital angular momentum during the evolution of neutron star recycling  \cite{bhattacharya1991formation, Zhu:2020zij}. 
Moreover, from the perspective of post-Newtonian expansion of the GW waveform,
align-ed spins are the dominant contribution from spins to the phasing of GWs and
the degeneracy between spins and symmetric mass ratio appears at 1.5 PN order, which could lead to large uncertainty in PE. 
Therefore, spin effects need to be considered in a more realistic CBC foreground subtraction scenario.
Secondly, we adopt the FIM to get a quick estimation. 
When degeneracy arises from two parameters in the waveform, the FIM  will be near-singular, which corresponds to an extremely high value of $c_{\rm{\Gamma}}$ \cite{Rodriguez:2011aa, Vallisneri:2007ev}. 
In the literature, those extreme events were arbitrarily discarded to ensure numerical precision when calculating the inverse of FIM \cite{Rodriguez:2011aa, Borhanian:2020ypi}. 
Here, this precision problem is solved with the help of the arbitrary-precision floating-point tool  \texttt{mpmath} \cite{mpmath}. 
Therefore, those extreme events are preserved in our simulation.
We discuss the effects of these extreme events on foreground subtraction. 

In this work, we first generate a population
of $10^5$ BBH events and $10^5$ BNS events up to a redshift of
$z\sim10$.\footnote{A more complete treatment can include neutron star--black
hole binaries as well~\cite{Zhu:2012xw}. Here, we use BNSs and BBHs to contrast
our results with that of \citet{Sachdev:2020bkk} and \citet{Zhou:2022nmt}.}  Then,
we employ an 11-dimensional PE (11-$d$ PE) for these BBH and BNS events using
the FIM method.  Comparing the results to those from the 9-dimensional PE (9-$d$
PE) by~\citet{Zhou:2022nmt}, we find that the residual of \reply{the} foreground becomes
even larger than the original background, which is primarily due to the
degeneracy between the spin parameters and the symmetric mass ratio.  These
results have significant implications for assessing the detectability of SGWB
from non-CBC origins for ground-based GW detectors.

This paper is arranged as follows.  In Sec.~\ref{Preliminary}, we introduce the
basics of the work, including the definition of the energy density spectrum of
GW events and the subtraction methods.  Also, we present our simulation methods
for generating BBH and BNS populations, \reply{the} configuration of \reply{an} XG detector network,
and the PE methods used in this work.  In Sec.~\ref{results}, we illustrate our
results and compare them with earlier results.  Some discussions are presented
in Sec.~\ref{Discussions}.

\section{Settings and Methods}
\label{Preliminary}

In this section, we present our settings for the calculation and the
consideration behind these settings.  We also explicitly spell out the details of
our methods in calculation.

\subsection{CBC population model}
\label{CBCpopulation}

Neglecting the tidal effects and detailed ringdown signals in BNSs, a generic
spinning, non-precessing, circular GW waveform is described by 11 parameters and
can be generated by the \texttt{IMRPhenomD} model \cite{Husa:2015iqa,
Khan:2015jqa}.  \citet{Zhou:2022nmt} considered 9 free parameters and fixed both
spins to zero.  In a PE process, it is more realistic to include the spin
effects.  Ideally, we shall consider generic spins, but here we restrain
ourselves to aligned spins only which contribute most significantly in the GW
phasing.  Therefore, we consider two more free parameters than
\citet{Zhou:2022nmt}, which are spins aligned or anti-aligned with the orbital
angular momentum.  The 11 free parameters we consider are 
\begin{align}
	\bm{\theta} = \big\{ m_1\,, m_2\ , D_L\ , \alpha \ , 
	\delta \ , \iota \ , \psi \ , \phi_c \ ,
	t_c \ , \chi_{1z} \ ,
	\chi_{2z} \big\},
\end{align}
where $m_1$ and $m_2$ are masses of the two components, and $\chi_{1z}$ and
$\chi_{2z}$ are spins paralleled with the orbital angular momentum.  

In our simulation, $10^5$ events are generated for BBHs and BNSs, respectively.
The population models are chosen \reply{as follows}.  Angle parameters such as
$\alpha$, $\psi$ and $\phi_{c}$ are drawn from a uniform distribution,
$\mathcal{U}[0, 2\pi)$, while $\cos \iota$ and $\cos \delta$ are drawn from
$\mathcal{U}[-1, 1]$. For the coalescence time, without losing generality, we set
$ t_\mathrm{c} = 0$, but still include it in the parameter estimation.

For the luminosity distance which is generated from redshift, we first consider
the local merger rate in the comoving coordinates,
\begin{equation}\label{Rm}
	R_{m}(z_m)= \int_{t_{d}^{\rm min}} ^{t_{d}^{\rm max}} R_{\rm sf}
	\Big\{z[t(z_m)-t_{d}]\Big\}p(t_{d}) \mathrm{d} t_{d}, 
\end{equation}
where $t(z_m)$ represents the cosmic time at merger at redshift $z_m$,
$R_{\mathrm{sf}}$ is the star-formation rate for binary systems whose details
can be found in Ref.~\cite{Vangioni:2014axa}. Additionally, $t_{d}$ denotes the
time delay between binary formation and merger, assumed to follow the
distribution \cite{Nakar:2007yr,Dominik:2012kk,Dominik:2013tma,
LIGOScientific:2016fpe, LIGOScientific:2017zlf, Dominik:2013tma,
Meacher:2015iua},
\begin{equation}\label{}
	p(t_d) \propto \frac{1}{t_d},  \  \ \quad \quad
	t_d^{\mathrm{min}}<t_d<t_d^{\mathrm{max}} ,
\end{equation}
where $t_d^{\mathrm{max}}$ is selected to be equal to the Hubble time, and 
\begin{equation}
	t_d^{\mathrm{min}} = \left\{
	\begin{array}{ll}
			20\, \mathrm{Myr} \,, & \quad\quad \mbox{for BNSs} \,, \\
			50\, \mathrm{Myr} \,, & \quad\quad \mbox{for BBHs} \,.
	\end{array}
		\right.
\end{equation}

Moreover, heavy BBHs are more likely to be formed in a low-metallicity environment \cite{LIGOScientific:2016fpe}.  When BBHs have the mass of at least one black hole greater than $30 M_\odot$, the star-formation rate in Eq.~(\ref{Rm}) needs to be modified into \cite{Callister:2016ewt}
\begin{equation}\label{}
	R_{\mathrm{BBH}}(z)\propto R_{\mathrm{sf}}(z)*F(z),
\end{equation}
where
\begin{equation}\label{}
	F(z)=\frac{\int_{-\infty}^{\log Z_{\odot}/2}\exp\left\{-2\left[\log Z-\overline{\log Z(z)}\right]^2\right\}d\log Z}{\int_{-\infty}^{\infty}\exp\left\{-2\left[\log Z-\overline{\log Z(z)}\right]^2\right\}d\log Z},
\end{equation}
with the metallicity of the Sun $Z_{\odot} = 0.02$, and the detail of $\overline{\log Z(z)}$ can be found in \citet{Callister:2016ewt}.

The distribution of redshift is obtained from the merger rate in the observer
frame \cite{Dominik:2013tma, Callister:2016ewt},
\begin{equation}\label{redshift distribution}
	R_{z}(z)=\frac{R_{m}(z)}{1+z}\frac{\mathrm{d}V_{c}(z)}{\mathrm{d}z} .
\end{equation}

For BBH mass parameters, we adopt the ``POWER LAW + PEAK'' mass model based on
the Gravitational Wave Transient Catalog 3 (GWTC-3) \cite{KAGRA:2021duu,
LIGOScientific:2018jsj}.  The primary mass follows a truncated power law
distribution, supplemented by a Gaussian component, 
\begin{widetext}
\begin{equation}\label{}
		P(m_1)	 \propto  S(m_1|m_{\mathrm{\mathrm{min}}},\delta_m) \times 
		\Big[ ( 1 - \lambda _ {
		\mathrm{peak}})P_{\mathrm{law}}(m_1|-\alpha,m_{\mathrm{\mathrm{max}}})
		+\lambda_{\mathrm{peak}}\mathcal{N}(\mu_m,\sigma_m)\Big]\,,
\end{equation}
\end{widetext}
where $\alpha =3.14$ and $m_{\mathrm{max}} = 86.85\, M_{\odot}$ for the
power-law  component, $\mu_m = 33.73\,  M_{\odot}$ and $\sigma_m = 3.36 \,
M_{\odot}$ for the Gaussian component, and $m_{\mathrm{min}} = 5.08 \,
M_{\odot}$ and $\delta_m = 4.83\, M_{\odot}$ for the smoothing function $S(\cdot
| \cdot)$ \cite{KAGRA:2021duu}. The weight parameter $\lambda_{\text{peak}} =
0.038$ is chosen as by~\citet{KAGRA:2021duu}.  The secondary mass population is
sampled from a conditional mass distribution over mass ratio $q=m_2/m_1$
\cite{KAGRA:2021duu, LIGOScientific:2018jsj},
\begin{equation}\label{}
	p(q) \propto q^{\gamma_{q}}S(m_{2}\mid m_{\min},\delta_{m}),
\end{equation}
where $\gamma_{q} = 1.08$.

For BNSs, the mass model is adopted from \citet{Farrow:2019xnc}. 
 The primary
mass $m_1$ is sampled from a double Gaussian distribution,
\begin{equation}\label{}
	P(m_{1})=\gamma_{\mathrm{NS}}\mathcal{N}\left(\mu_1, \sigma_1\right) +
	\left(1-\gamma_{\mathrm{NS}}\right)\mathcal{N}\left(\mu_2, \sigma_2\right),
\end{equation}
with $\gamma_{\mathrm{NS}} = 0.68$, $\mu_{1} = 1.34\, M_{\odot}$, $\sigma_{1} =
0.02\,M_{\odot}$, $\mu_{2}=1.47\,M_{\odot}$, and $\sigma_{2} = 0.15\,M_{\odot}$.
The secondary mass $m_2$ follows a uniform distribution,
$\mathcal{U}[1.14\,M_{\odot}, 1.46\,M_{\odot}]$. 

As for the spin parameters, we assume $\chi_{1z}$ and $\chi_{2z}$ to follow a
uniform distribution $\mathcal{U}\left[-1,1\right]$ for BBHs, and follow a
Gaussian distribution $\mathcal{N}\left(\mu_{\chi},\sigma_{\chi}\right)$ with
$\mu_{\chi} = 0$ and $\sigma_{\chi} = 0.05$ for BNSs \cite{Berti:2004bd}.

\subsection{Waveform reconstruction}
\label{waveform_reconstruction}

For large populations, the computational cost is expensive if one conducts a
full Bayesian PE for each event \cite{Lyu:2022gdr}.  Similarly to
\citet{Sachdev:2020bkk} and \citet{Zhou:2022nmt}, we adopt the FIM method to
recover parameters and their uncertainties.  We reconstruct waveforms from the
FIM results for both BBH and BNS events.

Assuming that the noise is stationary and Gaussian, under the linear-signal
approximation, the posterior distribution of GW parameters is \cite{Finn:1992wt,
Borhanian:2020ypi},
\begin{equation}\label{eq12}
	p(\boldsymbol{\theta})\sim
	\mathrm{e}^{-\frac12\Gamma_{ij}\Delta\theta_{i}\Delta\theta_{j}},
\end{equation}
where $\Gamma_{ij}$ is the FIM,
\begin{equation}
	\Gamma_{ij}\equiv \Big\langle\partial_{\theta_{i}}
	H(\boldsymbol{\theta};f),\partial_{\theta_{j}} H(\boldsymbol{\theta};f)
	\Big\rangle ,
\end{equation}
where $H(\boldsymbol{\theta};f)$ is the strain recorded in the detector and the
inner product for two quantities  $A(\boldsymbol{\theta};f)$ and
$B(\boldsymbol{\theta};f)$ is defined as,
\begin{equation}
	\langle A,
	B\rangle=2\int_{0}^{\infty}\mathrm{d}f\frac{A(\boldsymbol{\theta};f)
	B^{*}(\boldsymbol{\theta};f)+A^{*}(\boldsymbol{\theta};f)
	B(\boldsymbol{\theta};f)}{S_{n}(f)},
\end{equation}
where $S_n(f)$ is the one-side power spectrum density (PSD) for a specific
detector.

For a detected event, its matched-filter signal-to-noise ratio (SNR) is defined
as $\rho= \sqrt{\langle H ,H \rangle}$.  Then for a network with $N_d$
detectors, the corresponding SNR and FIM are respectively,
\begin{align}
	\rho_{\rm{net}} = \sqrt{\sum_{i=1}^{N_d}\rho_{i}^{2}}, \quad\quad
	 \Gamma_{\mathrm{net}} =\sum_{i=1}^{N_d}\Gamma_{i}.
\end{align}
We consider three XG detectors, including one CE with a 40-km arm length located
in Idaho, US, one CE with a 20-km arm length located in New South Wales,
Australia, and one ET with a triangular configuration located in Cascina, Italy.
This detector network corresponds to the fiducial scenario used by
\citet{Zhou:2022nmt}. 

After obtaining the FIM of 11 free parameters for each CBC event, we utilize
$\boldsymbol{\theta}^i_{\mathrm{tr}}$ and the covariance matrix
$\Sigma_{\mathrm{net}} \equiv  \Gamma_{\mathrm{net}}^{-1}$  to construct a
multivariate Gaussian distribution.  Subsequently, we employ this distribution
to randomly draw in the 11-$d$ parameter space to mimic the recovered GW
parameters, $\boldsymbol{\theta}^i_{\mathrm{rec}}$, which are employed to
generate the reconstructed GW waveforms, $
\tilde{h}_+(\boldsymbol{\theta}^i_{\mathrm{rec}};f)$ and $
\tilde{h}_\times(\boldsymbol{\theta}^i_{\mathrm{rec}};f)$.  We use the
\texttt{GWBENCH} package (version \texttt{0.7.1}) \cite{Borhanian:2020ypi} to
obtain the PSDs, generate the GW waveforms, and calculate SNRs and FIMs.  

It is worth noting that Eq.~(\ref{eq12}) is a good approximation for high SNR events~\cite{Vallisneri:2007ev}.  In principle, one should conduct a full Bayesian analysis for more accurate results, at least for those low SNR events where FIM is not applicable.
In this work,  we are more interested in a fast order-of-magnitude estimate for the CBC foreground subtraction.  Therefore, we adopt the FIM method for $10^5$ events as a compromise solution considering the accuracy and computational expense.

\subsection{Foreground subtraction methods}
\label{CBCsubtraction}

The dimensionless energy density spectrum  of GW, $\Omega_{\rm GW}$, is defined
as \cite{Phinney:2001di},
\begin{equation}\label{EnergySpectrum}
	\Omega_{\mathrm{GW}}(f) :=  \frac{f}{\rho_{c} c}F(f),
\end{equation}	
where $F(f)$ is the energy flux, $\rho_{c}=(3c^2H_0^2)/(8\pi G)$ is the critical
energy density, and $H_0$ is the Hubble constant.  The total flux of $N$ CBC
sources is given by \cite{Phinney:2001di},
\begin{equation}\label{MainFlux}
	F_{\mathrm{tot}} = \frac{\pi c^3}{2G}\frac{f^2}{T}\sum_{i=1}^{N}\left[
	\big|\tilde{h}^i_+(\boldsymbol{\theta}^i_{\mathrm{tr}};f) \big|^2 +
	\big|\tilde{h}^i_\times(\boldsymbol{\theta}^i_{\mathrm{tr}};f) \big|^2
	\right],
\end{equation}
where $ \tilde{h}^i_+(\boldsymbol{\theta}^i_{\mathrm{tr}};f)$ and
$\tilde{h}^i_\times(\boldsymbol{\theta}^i_{\mathrm{tr}};f) $ are plus and cross
modes of GWs from the $i$-th CBC event in the frequency domain, and $T$
corresponds to the total duration of the observation. 

To detect the SGWB from non-CBC origins with the XG ground-based GW detector network, we need to verify how well the CBC foreground can be subtracted.  If the subtraction performs well so that the residual spectrum $\Omega_{\rm res}$ is much smaller than the spectrum from a non-CBC origin SGWB, then it might be detected on the XG detector network.  During the subtraction, the $\Omega_{\rm res}$ comes from two parts:

\begin{equation}   	\Omega_{\mathrm{res}}=\Omega_{\mathrm{ns}}+\Omega_{\mathrm{err}},
\end{equation}
where $\Omega_{\mathrm{ns}}$ comes from those weak events which can not be resolved by the network, and $\Omega_{\mathrm{err}}$ comes from the imperfect subtraction of detected events.

Following \citet{Zhou:2022nmt}, a threshold SNR ${\rho}_{\rm thr}$, is
used to divide all the CBC events into two groups: those to be subtracted and
those not to be subtracted.  We denote $N_{\rm s}$ ($N_{\rm ns}$) as the number
of CBC events whose ${\rho}_{\rm net}$ is greater (less) than ${\rho}_{\rm
thr}$. Thus, the energy flux of the $N_{\rm ns}$ CBC events which are not to be subtracted is:
\begin{equation}
	\label{Fns} F_{\mathrm{ns}} = \frac{\pi
	c^3}{2G}\frac{f^2}{T}\sum_{i=1}^{N_{\rm ns}}\left[
	\big|\tilde{h}^i_+(\boldsymbol{\theta}^i_{\mathrm{tr}};f) \big|^2 +
	\big|\tilde{h}^i_\times(\boldsymbol{\theta}^i_{\mathrm{tr}};f) \big|^2
	\right],
\end{equation}

For the other $N_{\rm s}$ events which need to be subtracted, we first reconstruct the waveform $\tilde{h}^i_\times(\boldsymbol{\theta}^i_{\mathrm{tr}};f)$. This step is done by the FIM method mentioned in Sec.~\ref{waveform_reconstruction}.  Due to the existence of noise, there always is a mismatch between ${\theta}^i_{\mathrm{tr}}$ and ${\theta}^i_{\mathrm{rec}}$. 
Thus, a residual strain for each $N_{\rm s}$ event will be left in the data during the subtraction:
\begin{equation}
\label{residual}
	\delta \tilde{h}^i_{+/\times} =
	\tilde{h}^i_{+/\times}(\boldsymbol{\theta}^i_{\mathrm{tr}};f) -
	\tilde{h}^i_{+/\times}(\boldsymbol{\theta}^i_{\mathrm{rec}};f) \,.
\end{equation} 
This will finally contribute to the $\Omega_{\rm res}$ as imperfect subtraction part $\Omega_{\mathrm{err}}$.  The corresponding energy flux  of the residual strain is:  
\begin{equation}
	\label{Ferr}
		F_{\mathrm{err}} = \frac{\pi c^3}{2G}\frac{f^2}{T}\sum_{i=1}^{N_{\rm s}}
		\bigg[ \left| \delta \tilde{h}^i_{+} \right|^2   +
		\left| \delta \tilde{h}^i_{\times} \right|^2
		\bigg].
\end{equation}

In our simulation, the main work is to calculate the $\Omega_{\mathrm{ns}}$ and $\Omega_{\mathrm{err}}$ of our CBC population based on the FIM method.  There are also other ways to calculate $\Omega_{\mathrm{err}}$ by Eq.~(\ref{FerrSupp}) \cite{Pan:2023naq, Cutler:2005qq} in which one subtracts the reconstructed strain from the true strain recorded in the detector.  As shown in Appendix~\ref{SuppMethod}, the effects of subtraction of these two methods are in the same order, especially when the spin effects are considered.

\begin{figure*}[htbp] 
	\centering 
	\includegraphics[width=0.97\textwidth]{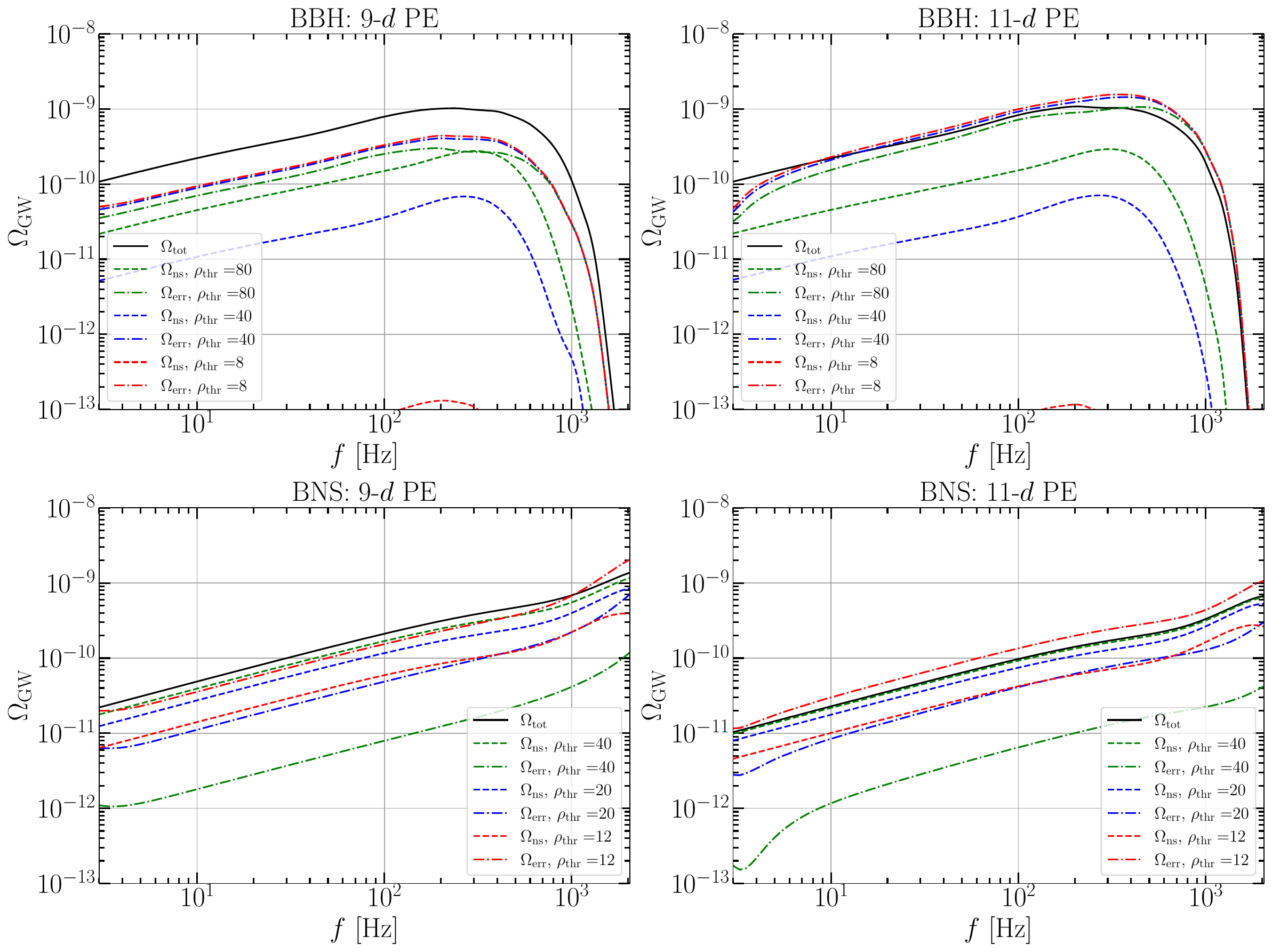} 
	\caption{Results from {\sc Treatment (I)} for 9-$d$ PE for BBHs (upper left)
	and BNSs (lower left), and 11-$d$ PE for BBHs (upper right) and BNSs (lower
	right).  Each subfigure shows the total GW energy spectrum
	$\Omega_{\mathrm{tot}}$ in black solid line and two components
	($\Omega_{\mathrm{ns}}$ and $\Omega_{\mathrm{err}}$) of the residual GW
	energy spectrum for different ${\rho}_{\rm thr}$.  $\Omega_{\mathrm{ns}}$
	(dash line) comes from the events that are not subtracted while
	$\Omega_{\mathrm{err}}$ (dash-dotted line) comes from the imperfect
	subtraction of the CBC foreground.  For a direct check, the two panels on
	the left reproduce the results of Fig. 2 in \citet{Zhou:2022nmt} for the
	\texttt{IMRPhenomD} waveform.}
	\label{strategy1} 
\end{figure*}
\begin{figure*}[htbp] 
	\centering 
	\includegraphics[width=0.97\textwidth]{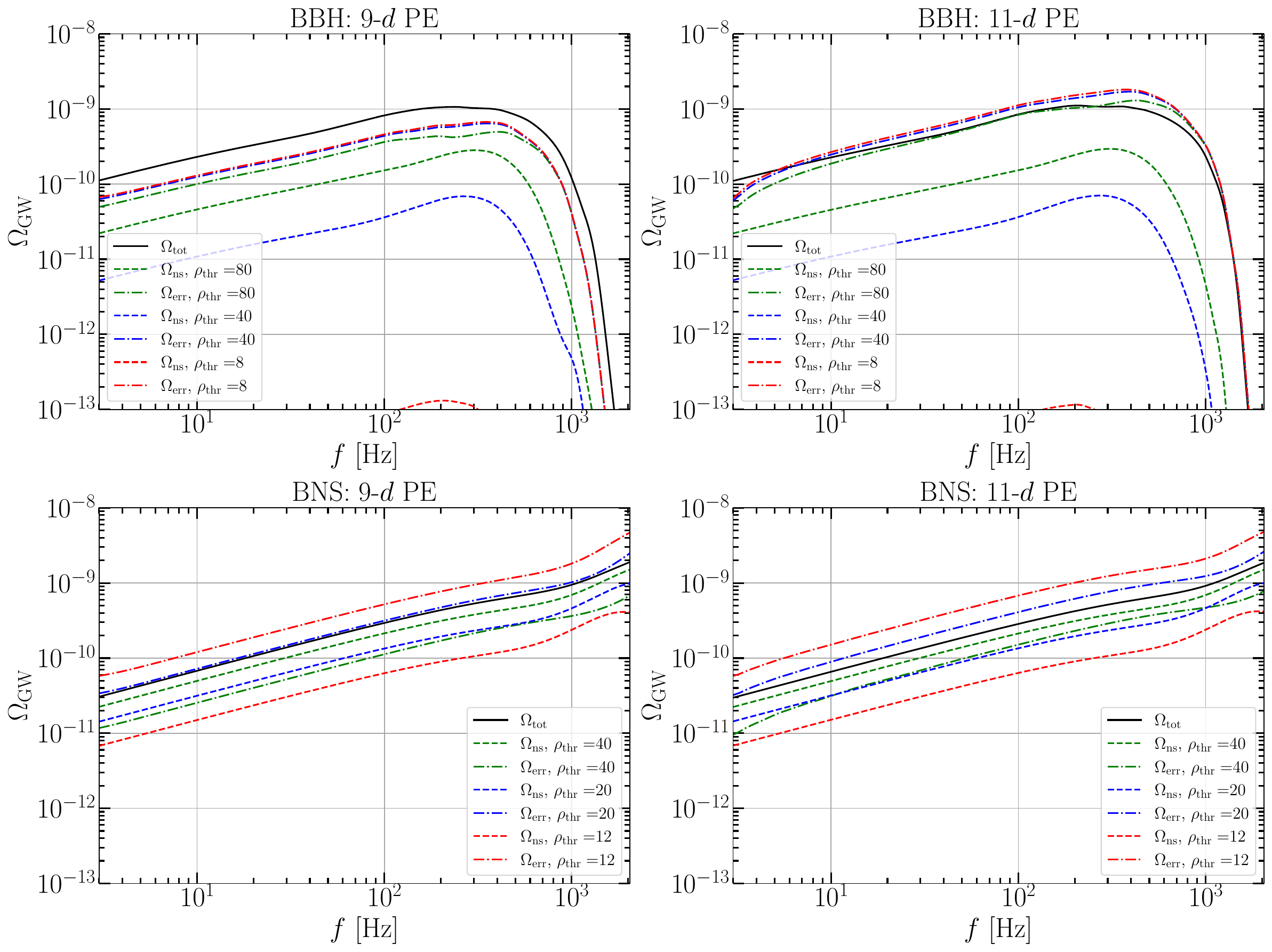} 
	\caption{Same as Fig.~\ref{strategy1}, but for {\sc Treatment (II)}.} 
	\label{strategy2} 
\end{figure*}

\section{Results}
\label{results}

When we try to obtain the covariance matrix through numerical inversion, there
is a condition number, $c_{\rm{\Gamma}}$, defined as the ratio between the largest eigenvalue and smallest one, 
which limits the accuracy of the numerical inversion of the FIM.  
As was pointed out by
\citet{Borhanian:2020ypi}, when using \texttt{GWBENCH 0.65}  to calculate the
covariance matrix, the inversion is not reliable when $c_{\rm{\Gamma}}$ exceeds
$10^{15}$. 
This is because that the commonly used \texttt{float64} format in \reply{the} computer only guarantees about 15 significant decimal digits. 
However, when two parameters are highly degenerate in the waveform, the FIM will be near-singular, which leads to a high value $c_{\rm{\Gamma}}$ \cite{Rodriguez:2011aa}. 
For some extreme events in our simulated population, the value of $c_{\rm{\Gamma}}$ will exceed $10^{15}$. While for the latest version \texttt{0.7.1},  \texttt{GWBENCH} uses
the \texttt{mpmath} \cite{mpmath} routine to ensure the accuracy of numerical inversion based
on the value of $c_{\rm{\Gamma}}$. 
Specifically, the floating-point numbers used in \texttt{mpmath} are represented as a 4-tuple ($sign$, $man$, $exp$, $bc$). 
The first three components are the $sign$, the $mantissa$ and the $exponent$, and they are normally used to save a number in the computer as $(-1)^{sign} \times man \times 2^{exp}$. 
The last component, the $bitcount$, is the newly defined parameter for saving the size of the absolute value of the mantissa in bits. 
The precision,  $prec$, for the float number depends on the maximum allowed $bc$ as $2^{prec}$ \cite{mpmath}. 
Compared to the default $prec=11$ in $float64$ number, the components in FIM saved in \texttt{mpmath} can be set to high numerical precision according to the value of $c_{\rm{\Gamma}}$.  Therefore, we keep all those events with a
large condition number in our simulations, rather than disregarding them
completely as mentioned in \cite{Rodriguez:2011aa}. 

Due to the different treatments of events with large condition number
$c_{\rm{\Gamma}}$, and for a consistency check with previous work
\cite{Zhou:2022nmt}, we employ two analysis treatments in the following
discussion: {\sc Treatment (I)} we subtract the events whose $\rho_{\rm net} >
\rho_{\rm thr}$ and $c_{\rm{\Gamma}}<10^{15}$; {\sc Treatment (II)} we subtract
the events as long as their $\rho_{\rm net} > \rho_{\rm thr}$.  {\sc Treatment
(I)} is consistent with the treatment in the previous version of
\texttt{GWBENCH} which was adopted by \citet{Zhou:2022nmt}, while {\sc Treatment
(II)} is consistent with the specifics in the version \texttt{0.7.1} of
\texttt{GWBENCH}.

We here consider four PE cases in our calculation: 
\begin{enumerate}[(i)]
	\item 9-$d$ PE for $10^5$ BBH events,
	\item 9-$d$ PE for $10^5$ BNS events, 
	\item 11-$d$ PE for $10^5$ BBH events, 
	\item 11-$d$ PE for $10^5$ BNS events. 
\end{enumerate}
The parameter configuration of the first two cases \reply{is} the same as in
\citet{Zhou:2022nmt} for validation and comparison reasons.  The results of
these four cases from {\sc Treatment (I)} are shown in Fig.~\ref{strategy1}, and
from {\sc Treatment (II)} are shown in Fig.~\ref{strategy2}.  In each figure,
the left panels show the results of 9-$d$ PE for BBH and BNS events, while the
right panels show the results for 11-$d$ PE cases.  We denote the spectrum
$\Omega_{\rm tot}$ with solid black line. Then, we choose three different
${\rho}_{\rm thr}$, i.e., ${\rho}_{\rm thr} = 8, 40, 80$ for BBH events, and
${\rho}_{\rm thr} = 12, 20, 40$ for BNS events.  For each ${\rho}_{\rm thr}$, we
denote the spectrum $\Omega_{\mathrm{ns}}$ with dash line and the spectrum
$\Omega_{\mathrm{err}}$ with dash-dotted line.  The left panels of both figures
reproduce well the results in \citet{Zhou:2022nmt}, and we find  similar
features that $\Omega_{\mathrm{err}}$ increases with ${\rho}_{\rm thr}$ while
$\Omega_{\mathrm{ns}}$ decreases with it.

However, the spectra $\Omega_{\mathrm{err}}$ in the left panels of
Fig.~\ref{strategy2} with {\sc Treatment (II)} are greater than that in
\citet{Zhou:2022nmt}, especially for the BNS 9-$d$ PE case.  This is due to the
contribution from events with $c_{\rm{\Gamma}}>10^{15}$ in our {\sc Treatment
(II)}.  Those events with high $c_{\rm{\Gamma}}$ values can lead to worse
subtraction results, thus contributing more to the spectrum
$\Omega_{\mathrm{err}}$, comparing to the events with  low $c_{\rm{\Gamma}}$
values.  To see it more clearly, we define two ratios for the $i$-th event: the
relative ratio $R_{\mathrm{rel}}$ and the absolute ratio  $R_{\mathrm{abs}}$,
\begin{equation}\label{eq:ratios}
		R_{\mathrm{rel}} = \frac{\delta h_i^2}{h_i^2},	 \quad\quad
		R_{\mathrm{abs}}  = \frac{\delta h_i^2}{\overline{h^2}} ,
\end{equation}
with
\begin{widetext}
\begin{align}\label{}
		\delta h_i^2 &= \big| \tilde{h}_+(\boldsymbol{\theta}^i_{\mathrm{tr}};f)
		-\tilde{h}_+(\boldsymbol{\theta}^i_{\mathrm{rec}};f) \big|^2 +\big|
		\tilde{h}_\times(\boldsymbol{\theta}^i_{\mathrm{tr}};f)
		-\tilde{h}_\times(\boldsymbol{\theta}^i_{\mathrm{rec}};f) \big|^2,	\\
		h_i^2  &=
		\big|\tilde{h}_+(\boldsymbol{\theta}^i_{\mathrm{tr}};f)
		\big|^2+\big|\tilde{h}_\times(\boldsymbol{\theta}^i_{\mathrm{tr}};f)\big|^2,\\
		\overline{h^2} &= \frac{1}{N}\sum_{i=1}^N h_i^2.
\end{align}
\end{widetext}
Notice that $\Omega_{\rm err}\propto \sum_{i=1}^{N_s} \delta h_i^2 $ and
$\Omega_{\rm tot}\propto \sum_{i=1}^N h_i^2$.  The value of $R_{\mathrm{rel}}$
represents the ratio of an event's contribution to  $\Omega_{\mathrm{err}}$ over
its contribution to $\Omega_{\mathrm{tot}}$.  Considering that the value of
$h_i^2$ varies from event to event, we employ $R_{\mathrm{abs}}$ to estimate the
ratio of an event's contribution to $\Omega_{\mathrm{err}}$ over the average
contribution to $\Omega_{\mathrm{tot}}$ across all $N$ CBC events.

\begin{figure*}[htbp] 
	\centering 
	\includegraphics[width=0.97\textwidth]{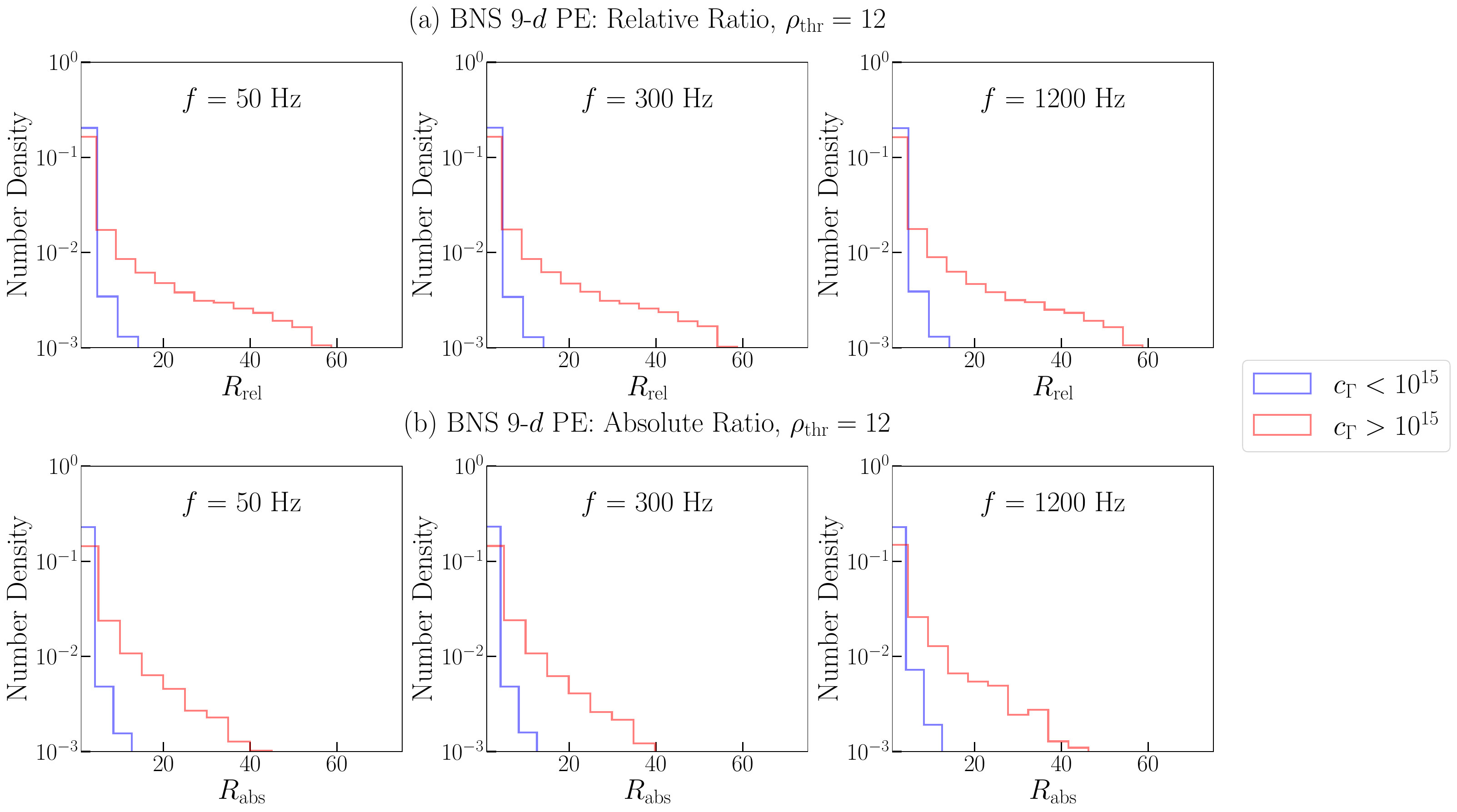} 
	\caption{Number density distribution of the relative ratio
	$R_{\mathrm{rel}}$ (upper panels) and the absolute ratio $R_{\mathrm{abs}}$
	(lower panels) for BNS 9-$d$ PE case with $\rho_{\rm thr}=12$. Plots for
	events with high/low $c_{\rm{\Gamma}}$ values (denoted with red/blue color)
	are shown separately.  Three frequency bins are chosen for illustration.  The
	distributions at 10, 100, 200, 400, 800 and 2000 Hz have similar features,
	which are not shown here.} 
	\label{BNScondnumVS} 
\end{figure*}

\begin{table}
	\renewcommand\arraystretch{1.3}
	\caption{Percentage of BBH and BNS events with $c_{\rm{\Gamma}} > 10^{15}$
	in our simulation.  For BBH cases we set $\rho_{\rm thr}=8$, while for BNS
	cases, we set $\rho_{\rm thr}=12$.
	}
	\label{condnum}
	\begin{ruledtabular}
		\begin{tabular}{ccc}
			& 9-$d$ PE & 11-$d$ PE\\
			\hline
			BBH  & 1.76\% & 4.32\%\\
			
			BNS  & 21.98\%  & 58.24\%\\
		\end{tabular}
	\end{ruledtabular}
\end{table}

We show the number density distribution of  $R_{\mathrm{rel}}$ and
$R_{\mathrm{abs}}$ for events with  high and low $c_{\rm{\Gamma}}$ values
separately in Fig.~\ref{BNScondnumVS} for the BNS 9-$d$ PE case with $\rho_{\rm
thr} = 12$.  Without loss of generality, three frequency bins are selected from
low to high for illustration.  We observe that events with high
$c_{\rm{\Gamma}}$ values (red line) are more concentrated at higher ratio values
than  events with low $c_{\rm{\Gamma}}$ values (blue line) for both
$R_{\mathrm{rel}} $ and $R_{\mathrm{abs}}$ at all chosen frequency bins.  This
indicates that events with high $c_{\rm{\Gamma}}$ values have a higher
probability of resulting in a worse subtraction than events with low
$c_{\rm{\Gamma}}$ values.  As shown in Table~\ref{condnum}, since there are
21.98\% events with high $c_{\rm{\Gamma}}$ values for BNS events with $\rho_{\rm
thr} = 12$, owing to the cumulative effects of these events, we observe a larger
$\Omega_{\mathrm{err}}$ in {\sc Treatment (II)} compared to that in {\sc
Treatment (I)}.  Similar results are obtained for the other three PE cases.
Thus, adopting {\sc Treatment (I)} rather than {\sc Treatment (II)} results in
an underestimation of $\Omega_{\rm err}$.

\begin{figure*}[htbp] 
	\centering 
	\includegraphics[width=0.95\textwidth]{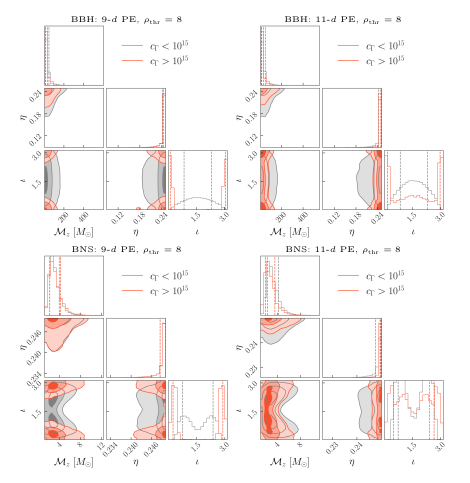} 
	\caption{Parameter distributions of events with high and low
	$c_{\rm{\Gamma}}$ values in four PE cases. In this figure, we set $\rho_{\rm
	thr} = 8$ for BBH cases and $\rho_{\rm thr} = 12$ for BNS cases. We show
	three parameters here: chirp mass in the observer frame $\mathcal{M}_z$,
	symmetric mass ratio $\eta$, and orbital inclination angle $\iota$. Red color
	denotes events with  $c_{\rm{\Gamma}}>10^{15}$, while gray color denotes
	events with $c_{\rm{\Gamma}} < 10^{15}$.} 
	\label{CornerPlot} 
\end{figure*}

Furthermore,  in Fig.~\ref{CornerPlot} we observe distinctive characteristics in
the parameter space for events with high $c_{\rm{\Gamma}}$ values (denoted with
red color), comparing to those with low $c_{\rm{\Gamma}}$ values (denoted with
gray color).  We set $\rho_{\rm thr} = 8$ for BBH cases and $\rho_{\rm thr} = 12$
for BNS cases.  The major difference is that the orbital inclination angle of events
with high  $c_{\rm{\Gamma}}$ values is likely to be distributed close to $0$ or
$\pi$.  It is not surprising, since there is strong degeneracy between the
parameter pairs, $\big\{ \iota, d_L \big\}$ and $\big\{\psi, \phi_c\big\}$, when
$\iota$ is close to $0$ or $\pi$ \cite{Cutler:1994ys, LIGOScientific:2013yzb,
Veitch:2014wba, Usman:2018imj}.  Besides, for events with high $c_{\rm{\Gamma}}$
values, the symmetric mass ratio concentrates much closer to 0.25, which means
that the two masses are nearly equal.

As discussed, the use of {\sc Treatment (I)} underestimates $\Omega_{\rm err}$,
so in the following we focus on the results of {\sc Treatment (II)}.  As is
shown in Fig.~\ref{strategy2}, when we incorporate aligned spins in PE, the
spectrum $\Omega_{\rm err}$ in the right panels grows significantly, comparing to
those of the 9-$d$ PE results in the left panels.  When comparing our new results in right panels in  Fig.~\ref{strategy2} with the results in \citet{Zhou:2022nmt}, we find that the $\Omega_{\mathrm{err}}$ will even surpass the total CBC foreground $\Omega_{\mathrm{tot}}$ for both BBH and BNS cases.  Specifically, for the BBH case, at 10 to 1000 Hz the $\Omega_{\mathrm{err}}$ for $\rho_{\rm thr}$ chosen as 8 and 40 will surpass the $\Omega_{\mathrm{tot}}$.  While for the BNS case,  the $\Omega_{\mathrm{err}}$ for $\rho_{\rm thr}$ chosen as 12 and 20 will surpass the $\Omega_{\mathrm{tot}}$ across the entire frequency band. More quantitatively, we compare the right panels of Fig.~\ref{strategy2} with \reply{the} left panels of Fig.~\ref{strategy1}, which closely follows the results from \citet{Zhou:2022nmt}.  For the BBH case, the $\Omega_{\mathrm{err}}$ for 11-$d$ PE is approximately 3 times greater than the 9-$d$ PE at 10 to 200 Hz, and approximately 5 to 10 times greater than the 9-$d$ PE case at 200 to 2000 Hz, for all three chosen $\rho_{\rm thr}$.  For the BNS case, we find that at 10 to 500 Hz, the  $\Omega_{\mathrm{err}}$ for 11-$d$ PE is approximately 4 times greater than the 9-$d$ PE for $\rho_{\rm thr}= 12$, approximately 8 times greater than the 9-$d$ PE for $\rho_{\rm thr}= 20$ and approximately 20 times greater than the 9-$d$ PE for $\rho_{\rm thr}= 40$.

\begin{figure*}[htbp] 
	\centering 
	\includegraphics[width=0.97\textwidth]{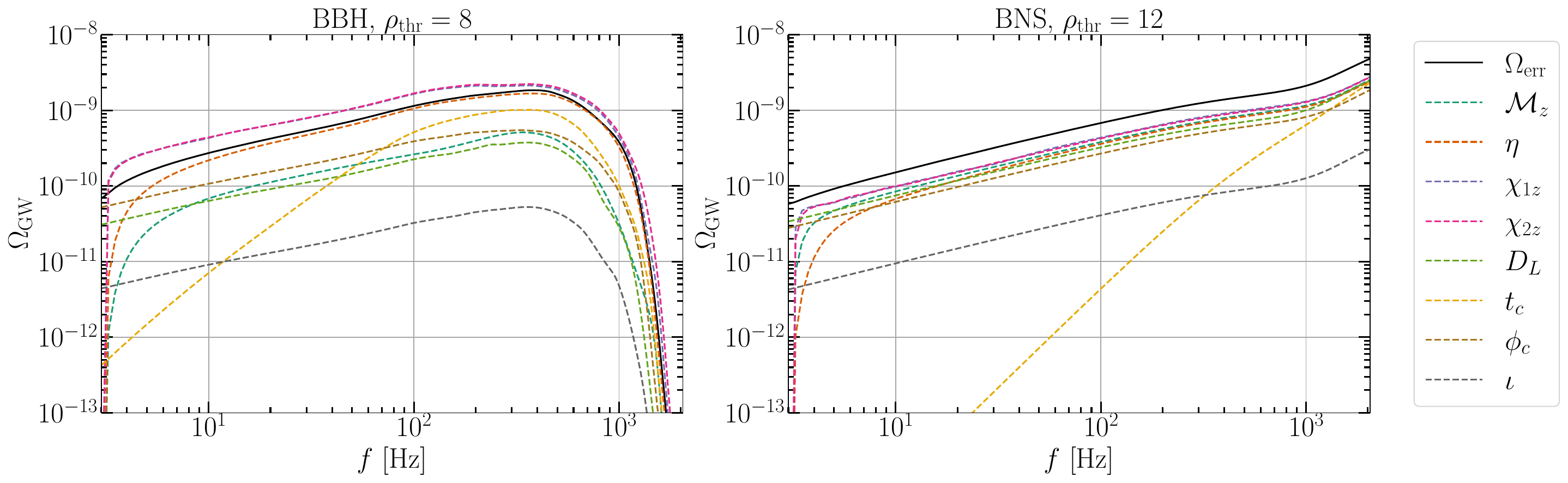} 
	\caption{Contribution from each parameter to $\Omega_{\mathrm{err}}$ for
	BBHs (left panel) and BNSs (right panel).  Each subfigure shows $\Omega_{\rm
	err}$ from 11-$d$ PE results (black solid line) and contribution from each
	parameter (dash line).  We choose  ${\rho}_{\rm thr}$ equal to 8 and 12 for
	BBHs and BNSs respectively.} 
	\label{WithSpinParaContribution} 
\end{figure*}

\begin{figure*}[htbp] 
	\centering 
	\includegraphics[width=0.97\textwidth]{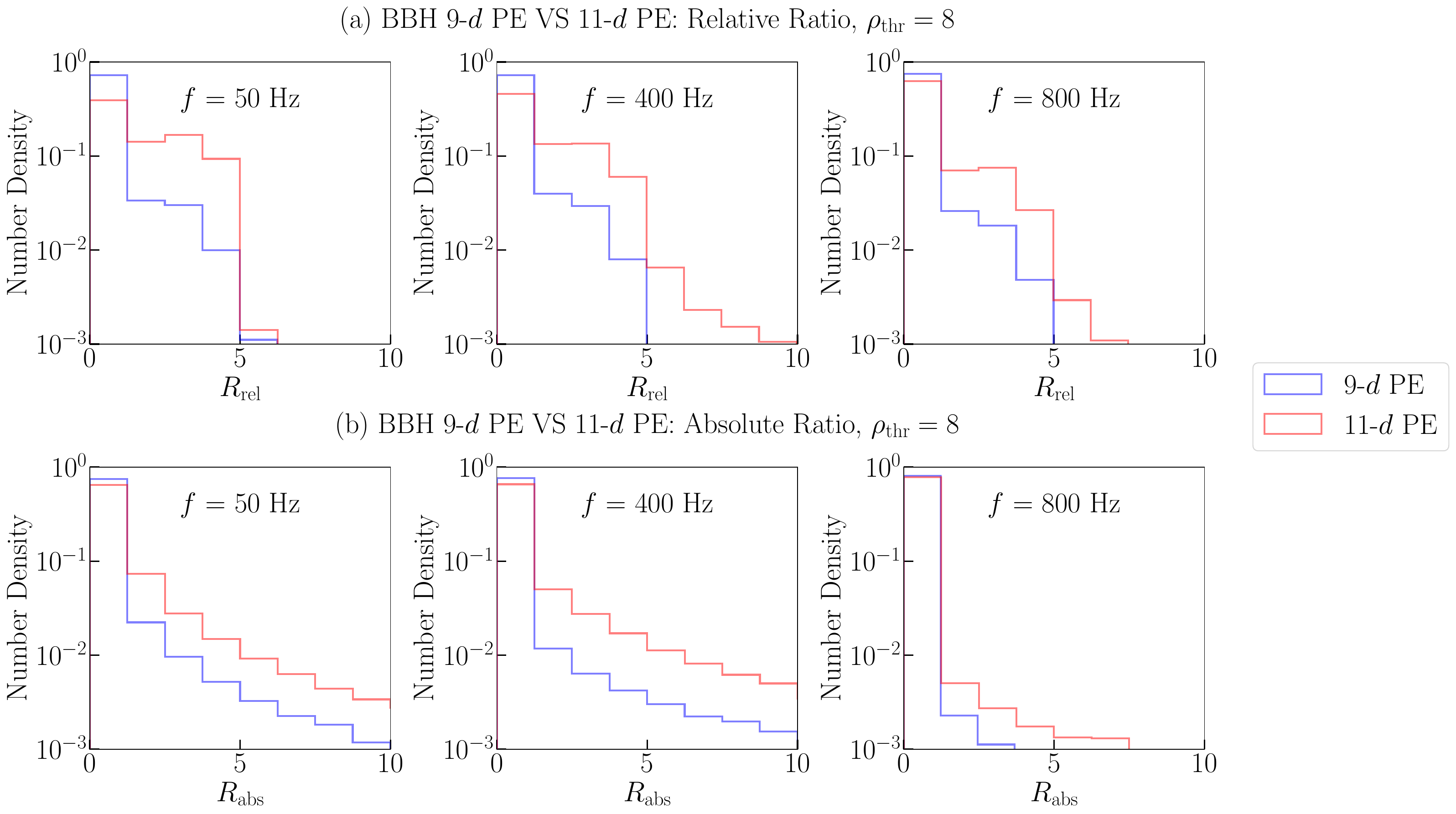} 
	\caption{The relative ratio $R_{\mathrm{rel}}$ (upper panels) and absolute
	ratio $R_{\mathrm{rel}}$ (lower panels) for 9-$d$ PE (blue line) and 11-$d$
	PE (red line) of BBHs. We set $\rho_{\rm thr}=8$ and choose 3 frequency bins
	to illustrate.  The BNS case has a similar feature in our simulation, which
	is not shown here. } 
	\label{BBH9VS11} 
\end{figure*}

The worse subtraction results mainly come from the
degeneracy between symmetric mass ratio and spins in the waveform model at the
inspiral stage \cite{Baird:2012cu, Hannam:2013uu, Ohme:2013nsa, Berry:2014jja,
Purrer:2015nkh, Farr:2015lna}.  As is shown in Table~\ref{corrcoef}, for the BBH
11-$d$ PE case, the absolute values of the correlation coefficients $C$ among
$\chi_{1z}$, $\chi_{2z}$ and $\eta$ exceed 0.99 for over 80\% of all events with
$\rho_{\rm thr} = 8$.  For the BNS 11-$d$ PE case, over 84\% of all events have
correlation coefficients exceeding 0.99 with $\rho_{\rm thr} = 12$.  The large
uncertainties due to the strong degeneracy will lead to a larger spectrum of
$\Omega_{\rm err}$.  To see it more clearly, we follow \citet{Zhou:2022nmt} to
estimate the contribution of each parameter to $\Omega_{\rm err}$. 
 For each
event, we reconstruct the waveform with the following choice of parameters. We
first choose the $k$-th parameter to be drawn from the 1-$d$ Gaussian
distribution with variance $\sigma_{\theta_k}$ (the $k$-th diagonal component of
the covariance matrix $\Sigma_{\rm net}$) when using the true value as its mean
$\mu_{\theta_k}$. Then, we set all the other parameters to be their true values.
Varying only one parameter and summing over all the subtracted events, we obtain
$\Omega_{\rm err}$ contributed from each parameter.  The results are shown in
Fig.~\ref{WithSpinParaContribution}, where we choose ${\rho}_{\rm thr}$ equal to
8 and 12 for BBHs and BNSs separately.  The contributions from $\chi_{1z}$,
$\chi_{2z}$ and $\eta$ dominate the spectrum $\Omega_{\mathrm{err}}$ even at
such high $\rho_{\rm thr}$ values.  For the BBH case, the contribution from
$\chi_{1z}$ and $\chi_{2z}$ even surpasses $\Omega_{\mathrm{err}}$ from the
11-$d$ PE results.  Meanwhile, for the BNS case, there are also subdominant
contributions from $D_L$, $\phi_c$, and $\mathcal{M}_z$.  Similar results were
also found by \citet{Zhou:2022nmt}.  As a result, the discrepancy between
parameters becomes more pronounced in 11-$d$ PE cases than in the 9-$d$ PE
cases.  Both the relative ratio $R_{\mathrm{rel}}$ and the absolute ratio
$R_{\mathrm{abs}}$ have grown significantly, comparing to those in the 9-$d$ PE
cases with $\rho_{\rm thr} = 8$, which is illustrated in Fig.~\ref{BBH9VS11} for
BBHs.  Hence, we observe a larger spectrum $\Omega_{\rm err}$.

\begin{table}
	\renewcommand\arraystretch{1.3}
	\caption{Percentage of BBH and BNS events with the absolute values of
	correlation coefficients among spins and symmetric mass ratio greater than
	0.99 for 11-$d$ PE cases.  For the BBH case we set $\rho_{\rm thr}=8$, and
	for the BNS case, $\rho_{\rm thr}=12$.
	}
	\label{corrcoef}
	\begin{ruledtabular}
		\begin{tabular}{cccc}
			& $\big|C_{\chi_{1z}, \chi_{2z}}\big|>0.99$ & $\big|C_{\eta,
			\chi_{1z}}\big|>0.99$ & $\big|C_{\eta, \chi_{2z}}\big|>0.99$ \\
			\hline
			BBH  & 98.80\% & 81.05\% & 79.98\%\\
			
			BNS  & 99.96\%  & 84.17\% & 84.05\%\\
		\end{tabular}
	\end{ruledtabular}
\end{table}

As was mentioned by \citet{Zhou:2022nmt}, there exists \reply{an} optimal $\rho_{\rm
thr}$ to minimize the spectrum $\Omega_{\mathrm{res}}$ for the 9-$d$ PE cases. 
For the 11-$d$ PE cases, following their approach, at almost all frequency bands
we find an optimal ${\rho}_{\rm thr} = 373$ for BBHs and an optimal ${\rho}_{\rm
thr}=200 $ for BNSs to minimize the spectrum $\Omega_{\mathrm{res}}$.  This
implies that in our simulated populations, only 612 BBH events and 7 BNS events
are to be subtracted in the XG GW detector network, which is unrealistic.
Therefore, a better approach is pressingly needed to deal with this issue.

\section{Summary}
\label{Discussions}

Considering an XG detector network which includes one ET and two CEs, we estimate how well the CBC foreground can be subtracted by the FIM method, which can be used to estimate the possibility of detecting SGWB from non-CBC origins in future studies.  To subtract the $i$-th GW event whose true
waveform is $\tilde{h}^i_{+/\times}(\boldsymbol{\theta}^i_{\mathrm{tr}};f)$, we
first use FIM method to approximately get the posterior distribution of the
parameters.  Then, we randomly draw parameters
$\boldsymbol{\theta}^i_{\mathrm{rec}}$ from this distribution to reconstruct the
waveform, $\tilde{h}^i_{+/ \times}(\boldsymbol{\theta}^i_{\mathrm{rec}};f)$.
After subtracting the reconstructed waveform, there is some residual $\delta
\tilde{h}^i_{+/\times}$  in the data [see Eq.~(\ref{residual})].  The summation
over all the subtracted events we obtain the spectrum $\Omega_{\rm err}$, which
corresponds to the energy density brought by the imperfect foreground
subtraction.  Then, $\Omega_{\rm err}$, combined with $\Omega_{\rm ns}$, which
is the spectrum of the GW events that are not subtracted because of low SNRs,
forms $\Omega_{\rm res}$.  In reality, we want to minimize $\Omega_{\rm res}$ so
as to detect SGWB from non-CBC origins.

As an extension of the previous studies by \citet{Sachdev:2020bkk} and
\citet{Zhou:2022nmt}, two more new features are considered during the subtraction.  Firstly, we include spin parameters in PE, in other words, we adopt
an 11-$d$ PE using the FIM for the CBC events to be subtracted.  For a realistic
consideration, we generate $10^5$ BBH and BNS events based on the latest
population models provided by the LVK collaboration \cite{KAGRA:2021duu} and
consider different treatments for subtraction of events that have large
condition numbers when inverting the FIM.  Secondly, we discuss the effects of the extreme events with a high value of $c_{\rm{\Gamma}}$.

When we do the subtraction, we first set a threshold SNR $\rho_{\rm thr}$.  For those low SNR events with
$\rho_{\rm net}<\rho_{\rm thr}$, we do not subtract them since the PE
uncertainties of these events are too large, and some events are even unsolvable
if $\rho_{\rm net} \lesssim 8$.  However, there is still no guarantee to well
reconstruct the true waveform for an event with a large SNR.  Sometimes, there
can be strong degeneracy between some parameters in the waveform model, which
leads to a large deviation between the reconstructed waveform and the true
waveform.  When the degeneracy between the parameters is strong, the condition
number $c_{\rm{\Gamma}}$ of FIM can be very large.  We propose two treatments,
for {\sc Treatment (I)}, we subtract the events whose $\rho_{\rm net} >\rho_{\rm
thr}$ and $c_{\rm{\Gamma}} <10^{15}$, and for {\sc Treatment (II)}, we subtract
all the events as long as $\rho_{\rm net} >\rho_{\rm thr}$.  Comparing the
results of {\sc Treatment (I)} in Fig.~\ref{strategy1} and {\sc Treatment (II)}
in Fig.~\ref{strategy2}, we find significant contribution to $\Omega_{\rm err}$
from events with large $c_{\rm{\Gamma}}$.  We verify it by calculating the
distribution of  $R_{\rm rel}$ and  $R_{\rm abs}$ [see Eq.~(\ref{eq:ratios})],
as shown in Fig.~\ref{BNScondnumVS}.  To conclude, the early study
underestimated $\Omega_{\rm err}$ when discarding events with large
$c_{\rm{\Gamma}}$.  To be more realistic, we include these events in our
calculation.  We also study the characteristics of the distribution of
parameters when $\rho_{\rm net}>\rho_{\rm thr}$ and $c_{\rm{\Gamma}}>10^{15}$. 
The orbital inclination angle $\iota$ is much more likely to distribute around $0$ or
$\pi$ for these events (see Fig.~\ref{CornerPlot}), which leads to degeneracy
between $\iota$ and $D_L$, and $\psi$ and $\phi_c$.  Besides, the symmetric mass
ratio is more likely to be closer to 0.25 for events with high $c_{\rm{\Gamma}}$
values.  By introducing higher order modes in the waveform model, we may break
the degeneracy between $\iota$ and $D_L$, and $\psi$ and $\phi_c$ to some extent
\cite{Lasky:2016knh, Payne:2019wmy, Zhang:2023ceh, Gong:2023ecg}, especially for
the events with asymmetric masses
\cite{LIGOScientific:2020stg, LIGOScientific:2020zkf} or high masses
\cite{Chatziioannou:2019dsz,LIGOScientific:2020ufj}.  The uncertainty in PE for
events with spins can also be reduced by including non-quadrupole modes
\cite{Varma:2014jxa,CalderonBustillo:2015lrt,Varma:2016dnf}.  From this
perspective, we expect to get a more optimistic result of the foreground
subtraction by using a waveform model including higher order modes in future
studies.

We compare our results with those obtained by \citet{Zhou:2022nmt}, where a
9-$d$ PE was adopted.  After including the aligned spins, the degeneracy between
parameters becomes worse, especially between the spin parameters and the
symmetric mass ratio.  As is shown in Fig.~\ref{WithSpinParaContribution}, the
effects from $\chi_{1z}$, $\chi_{2z}$ and $\eta$ surpasses that from $\phi_c$
which dominates in the 9-$d$ PE \cite{Zhou:2022nmt}.  The degeneracy increases
the uncertainty when performing PE and results in unexpectedly large
$\Omega_{\rm res}$, which is even larger than $\Omega_{\rm tot}$. 

In this work, we only consider the uncertainty of PE brought by the noise.  When
the error from inaccurate waveform modeling cannot be neglected
\cite{Cutler:2007mi,Gamba:2020wgg,Purrer:2019jcp,Hu:2022bji}, it also needs to
be discussed quantitatively \reply{in future studies}.  Last but not least, we have
assumed that GW signals can be identified and then subtracted one by one in the
literature.  However, it seems very optimistic for XG detectors since there
can be plenty of GW signals overlapping with each other, making PE more
difficult \cite{Regimbau:2012ir, Meacher:2015rex, Samajdar:2021egv,
Pizzati:2021apa, Relton:2021cax, Wang:2023ldq,Dang:2023xkj}.  We have to take
into account the effects of overlapping between signals in future studies. 

\acknowledgments 
We thank Zhenwei Lyu, Xing-Jiang Zhu, Zhen Pan, Huan Yang and the
anonymous referee for helpful comments.  This work was supported by the
Beijing Natural Science Foundation (1242018), the National Natural Science
Foundation of China (11975027, 11991053,  11721303), the China Postdoctoral
Science Foundation (2021TQ0018), the National SKA Program of China
(2020SKA0120300), the Max Planck Partner Group Program funded by the Max Planck
Society, and the High-Performance Computing Platform of Peking University.

\appendix

\section{Supplementary waveform subtraction method and results}
\label{SuppMethod}

In the main context, following the methods from \citet{Sachdev:2020bkk} and \citet{Zhou:2022nmt}, we obtain the residual for each event by subtracting the reconstructed plus and cross GW polarization waveforms from the true waveforms. 
Additionally, we introduce a supplementary subtraction method from \citet{Cutler:2005qq} and \citet{Pan:2023naq}.  In this method, the primary residual is obtained by subtracting the reconstructed strain from the true strain recorded in the detector. \reply{We still use the same parameter basis as the main context rather than the re-parametrization basis in \citet{Pan:2023naq}.} The strain signal for a specific event is:
\begin{equation}
H(\boldsymbol{\theta};f)=F_{+}(\alpha,\delta,\psi)\tilde{h}_{+}(f)+F_{\times}(\alpha,\delta,\psi)\tilde{h}_{\times}(f),
\end{equation}
where $F_{+}$ and $F_{\times}$ are the antenna pattern functions of the detector. 
Then, the energy flux can be expressed in terms of GW strain signal as \cite{Pan:2023naq}:
\begin{equation}
F_{\rm tot}(f)=\frac{2}{\langle F_+^2\rangle+\langle F_\times^2\rangle}\frac{\pi c^3}{2G}\frac{f^2}{T}\sum_{i=1}^N|H^i(f)|^2,
\end{equation}
where $\langle F_+^2\rangle$ and $\langle F_\times^2\rangle$ are the angle-averaged antenna pattern functions. Following the discussion in \citet{Pan:2023naq}, a CE with a 40-km arm located in Idaho, US, was considered as the reference detector. For such a L-shape interferometer,  $\langle F_+^2\rangle = \langle F_\times^2\rangle =\frac{1}{5
}$. When it comes to the foreground subtraction, the residual for each event is:

\begin{equation}
\delta H^i(f)=H^i(\boldsymbol{\theta}^i_{\mathrm{tr}}, f)-H^i(\boldsymbol{\theta}^i_{\mathrm{rec}}, f),
\end{equation}
where the recovered parameters $\boldsymbol{\theta}^i_{\mathrm{rec}}$ are obtained by the FIM methods using {\sc Treatment (II)}. Then, the energy flux due to imperfect subtraction is:
\begin{equation}
\label{FerrSupp}
    F_{\rm err}(f)=\frac{2}{\langle F_+^2\rangle+\langle F_\times^2\rangle}\frac{\pi c^3}{2G}\frac{f^2}{T}\sum_{i=1}^{N_s}|\delta H^i(f)|^2
\end{equation}
and the flux for $N_{\rm ns}$ unsubtracted events is: 
\begin{equation}
     F_{\mathrm{ns}} = \frac{\pi
	c^3}{2G}\frac{f^2}{T}\sum_{i=1}^{N_{\rm ns}}|H^i(f)|^2.
\end{equation}

\begin{figure*}[htbp] 
	\centering 
	\includegraphics[width=0.97\textwidth]{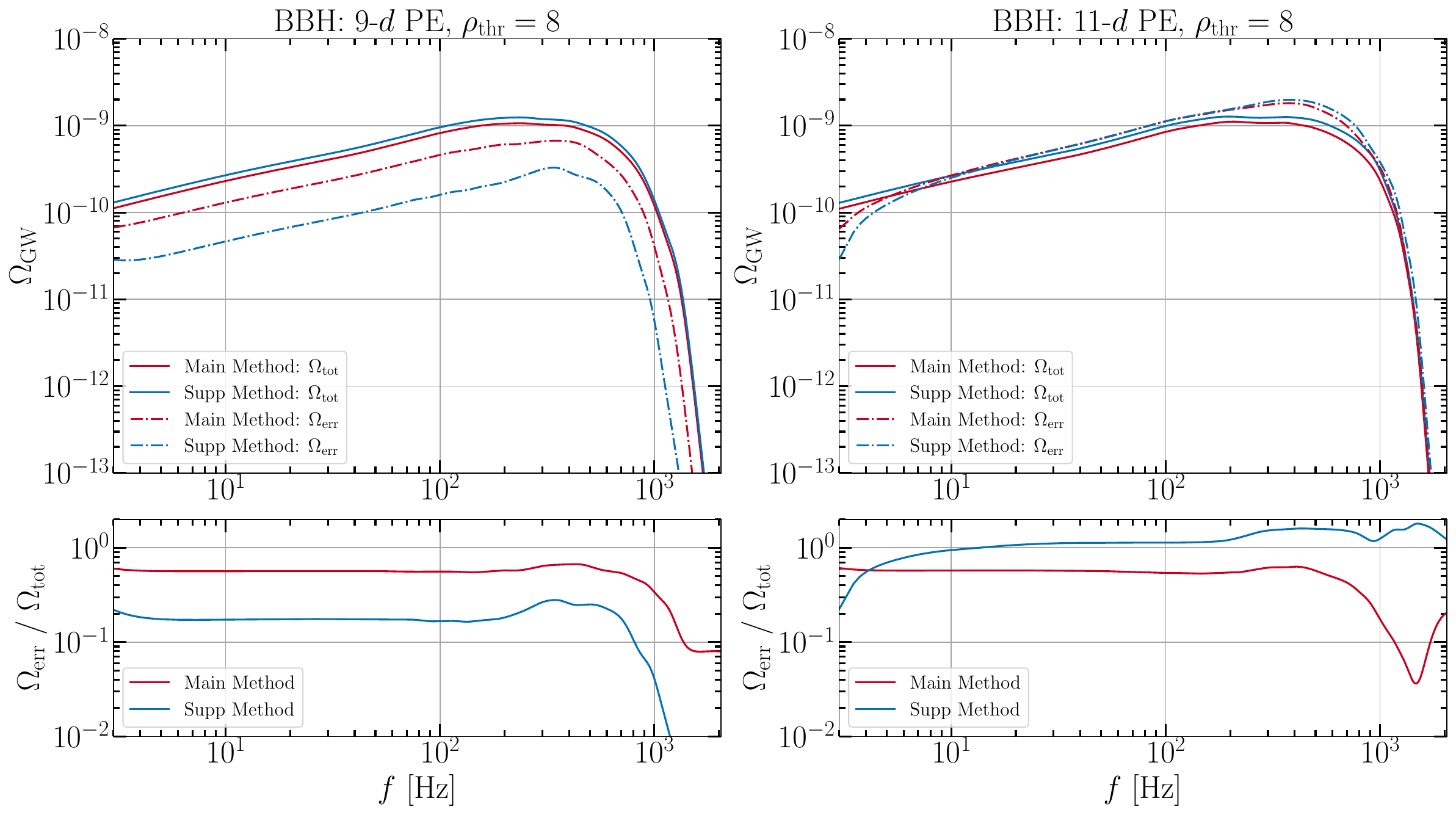} 
	\caption{The foreground subtraction results are presented for the main method (from main text) with red color and the supplementary method (from this Appendix) with blue color.  The left panels show the results for 9-$d$ PE, while the right panels show the results for 11-$d$ PE. In the top panels, each subfigure shows the $\Omega_{\mathrm{tot}}$ with solid
  line and $\Omega_{\mathrm{err}}$ with $\rho_{\rm thr}=8$ with dash-dotted line. In the bottom panels, each subfigure shows the results of $\Omega_{\mathrm{err}}$ against  $\Omega_{\mathrm{tot}}$.} 
	\label{AppendixFig}
\end{figure*}

For an illustration, we consider the same BBH population as in the main text for calculating the $\Omega_{\mathrm{ns}}$ and $\Omega_{\mathrm{err}}$. 
The results are shown in Fig.~\ref{AppendixFig}, in which we plot the $\Omega_{\mathrm{ns}}$ and $\Omega_{\mathrm{err}}$ for the main method as red color and for the supplementary method as blue color. 
Both the 9-$d$ PE case and the 11-$d$ PE case are considered. 
We also show the comparison of these two methods in the bottom panels for $\Omega_{\mathrm{err}}$ against  $\Omega_{\mathrm{tot}}$. 
Since we want a better foreground subtraction, a smaller value is preferred.
There are slight differences for the $\Omega_{\mathrm{tot}}$ in two methods because of the different treatments of $\alpha,\delta,\psi$ in Eq.~(\ref{FerrSupp}) and Eq.~(\ref{MainFlux}).
We find that for the 9-$d$ PE case, the supplementary subtraction method gives more positive results than the main method for approximately 3 times at 10 to 1000 Hz. 
For the 11-$d$ case, the main method gives more positive results than the supplementary subtraction method for approximately 2 times around 10 to 1000 Hz.  Meanwhile,
the supplementary subtraction method for 11-$d$ PE case also shows that the $\Omega_{\mathrm{err}}$ surpasses the $\Omega_{\mathrm{tot}}$ at 10 to 1000 Hz, which is consistent with our results in the main text. 
Moreover, since in the results of 11-$d$ PE case, the main method shows a slightly better subtraction effect, we consider the results in the main text quite complementary to the methods in other works.

\bibliographystyle{apsrev4-1}
\bibliography{subtractionref}

\end{document}